\documentclass[useAMS,usenatbib]{mn2e}

\usepackage{times}
\usepackage{graphicx}

\newcommand{\beq}{\begin{equation}}
\newcommand{\eeq}{\end{equation}}
\newcommand{\lab}{\label}

\def\gs{\mathrel{\lower0.6ex\hbox{$\buildrel {\textstyle >}\over{\scriptstyle \sim}$}}}
\def\ls{\mathrel{\lower0.6ex\hbox{$\buildrel {\textstyle <}\over{\scriptstyle \sim}$}}}

\begin{document}

\title[Kinematic effect in GL by galaxy clusters]{Kinematic effect in gravitational lensing by clusters of galaxies}
\author[M. Sereno]{M. Sereno\thanks{E-mail:sereno@physik.unizh.ch} 
\\
Institut f\"{u}r Theoretische Physik, Universit\"{a}t Z\"{u}rich,
Winterthurerstrasse 190, CH-8057 Z\"{u}rich , Switzerland
}


\maketitle

\begin{abstract}
Gravitational lensing provides an efficient tool for the investigation of matter structures, independent of the dynamical or hydrostatic equilibrium properties of the deflecting system. However, it depends on the kinematic status. In fact, either a translational motion or a coherent rotation of the mass distribution can affect the lensing properties. Here, light deflection by galaxy clusters in motion is considered. Even if gravitational lensing mass measurements of galaxy clusters are regarded as very reliable estimates, the kinematic effect should be considered. A typical peculiar motion with respect to the Hubble flow brings about a systematic error $\ls 0.3\%$, independent of the mass of the cluster. On the other hand, the effect of the spin increases with the total mass. For cluster masses $\sim 10^{15}M_\odot$, the effect of the gravitomagnetic term is $\ls 0.04\%$ on strong lensing estimates and $\ls 0.5\%$ in the weak lensing analyses. The total kinematic effect on the mass estimate is then $\ls 1\%$, which is negligible in current statistical studies. In the weak lensing regime, the rotation imprints a typical angular modulation in the tangential shear distortion. This would allow in principle a detection of the gravitomagnetic field and a direct measurement of the angular velocity of the cluster but the required background source densities are well beyond current tecnological capabilities.
\end{abstract}

\begin{keywords}
galaxies: clusters: general -- gravitational lensing -- cosmology: observations
\end{keywords}

\section{Introduction}

Clusters of galaxies are the biggest things whose masses can be reliably measured. The measurements of their properties are prerequisites to understand the structure in the universe on a very large scale and to investigate processes associated with galaxy formation \citep{voi05}. Investigations are often performed using rather strong assumptions. Mass estimates based on X-ray observations are routinely obtained through the hydrostatic equilibrium equation. Such measurements can be quite accurate if the temperature profile is well reconstructed from the projected measured one \citep{nag+al07} but they can be biased low by 5-10\% through the virial region primarly due to neglecting the unknown pressure support provided by gas bulk motion \citep{ras+al06,nag+al07}. The complex thermal structure of the emitting plasma can also bias towards lower values \citep{ras+al06}. The mass of a steady cluster can be also inferred putting the observed velocity dispersion through the virial theorem \citep[and references therein]{voi05}. However, assumptions must be made on the degree of anisotropy to relate the projected velocity dispersion to the intrinsic components. A boundary pressure term can also alter the viral relation. 

The hypotheses of either hydrostatic or dynamical equilibrium might be not suitable in many systems. Cluster of galaxies are the latest object to form in a hierarchical cold dark matter scenario and many of them are likely to be still in the process of formation. Gravitational lensing offers a theoretically less demanding alternative approach independent of the physical state and nature of the matter. In fact, the mass measurement is reliable even in merging clusters. In general, two shortcomings are recognised as affecting gravitational lensing estimates. First, projection effects, as happens also for other methods, limit the accuracy. In fact, lensing measures the mass of all the structures superimposed to the cluster \citep{met+al99}. Second, on a more theoretical ground, the steepness degeneracy makes the lensing properties invariant for a local rescaling \citep{sah+al06}. However, this can be broken having a range of source redshifts, with a very large field of view or having number counts of lensed images. 

Lensing methods do not rely on equilibrium hypotheses but, even though the implicit assumption of a static mass distribution is usually made, kinematics actually affects the lensing properties of a mass distribution \citep[ans references therein]{ser02}. Either the peculiar motion of the deflector with respect to the Hubble flow or a coherent rotation of the matter halo bring in corrections to gravitational lensing. The nature of the two effects is substantially different \citep{ser05b}. The effect due to a translational motion comes from the local Lorentz invariance and from the existence of the Newtonian (gravitoelectric) field \citep[and references therein]{fri03,ser05b}. On the other hand, the mass current induced by a non null angular momentum induces a gravitomagnetic field, which is peculiar of general relativity and other metric theories of gravity, and the related dragging of inertial frames \citep[and references therein]{ser02,ser03b}. The lens motion can affect observations on very different scale-lengths: Galactic microlensing \citep{ser03a}, black hole lensing both in the weak \citep{ser03a,se+de06} and strong deflection limit \citep{boz+al05,boz+al06}, time-delays \citep{ser05a} and deflection angles \citep{cap+al03} in galaxy-quasar lensing can show sizable signatures of either spin or translational motion. Dark matter currents in the large-scale structure also affect the weak-lensing power spectrum, even if corrections are negligible at currently accessible scales \citep{sc+ba06}.

In this paper, I will discuss the effect of motion of galaxy clusters on gravitational lensing. The translational and rotational motions of galaxy clusters are strictly related to the formation and evolution of large scale structure. Peculiar velocities \citep{ba+oh96,mas+al06} and spins \citep{bet+al07,go+ye07,hw+le07} can be sizable and their effect deserves attention. I will consider the peculiar lensing signatures imprinted by the motion of the cluster and how the kinematics of the deflector affects cluster mass estimators, in the weak as well as in the strong lensing regime. The paper is organised as follows. In Section~\ref{sec:rSIS}, the properties of a model of rotating and translating lens are reviewed. Section~\ref{sec:pecu} and  Section~\ref{sec:spin} are devoted to the effect of peculiar motions and angular momentum, respectively. Finally, Section~\ref{sec:conc} contains some concluding remarks. Throughout this paper the reference cosmological model is, unless otherwise stated, a flat model of universe with a cosmological constant ($\Omega_\mathrm{M}=0.3$, $\Omega_{\Lambda}=0.7$) and $H_0=100 h \mathrm{km\ s}^{-1} \mathrm{Mpc}^{-1}$.

\section{Rotating isothermal sphere}
\label{sec:rSIS}

Many of the properties of galaxy clusters can be understood using a very simple model in which the matter distribution is treated as a singular isothermal sphere (SIS),
\beq
\lab{sis7}
\rho (r) = \frac{\sigma_\mathrm{v}^2}{2 \pi G r^2},
\eeq
where $r$ is the radial distance, $\sigma_\mathrm{v}$ is the velocity dispersion and $G$ the gravitational constant. This model predicts quite correctly many self-similar features and scaling relations \citep{voi05}. Since the total mass of a SIS is divergent, a cut-off radius much larger than the relevant length scale which characterises the lensing phenomenon must be introduced. Based on the spherical collapse model, the limiting radius can be defined to be $r_\Delta$, the radius within which the mean mass density is $\Delta$ times the critical density of the universe $\rho_\mathrm{cr}=3H(z)/(8 \pi G)$ where $H(z) =H_0 \sqrt{\Omega_\mathrm{M}(1+z)^3 +(1-\Omega_\mathrm{M})}$ is the time dependent Hubble parameter. For a SIS at redshift $z_{\rm d}$, it is \citep{mo+al98}
\beq
r_\Delta = \frac{2 \sigma_\mathrm{v} }{ \sqrt{\Delta} H(z_{\rm d})}.
\eeq
No single definition of mass overdensity is best for all applications regarding galaxy clusters \citep{voi05}. A useful approximation is based on the spherical top-hat model. For our reference $\Lambda$CDM model, $\Delta \sim 155.5$ at $z \simeq 0.3 $ \citep{br+no98}. Then, a halo with $\sigma_\mathrm{v} \sim 800~\mathrm{km~s}^{-1}$ at $z_\mathrm{d}=0.3$ has a virial radius of $\sim 1.1~\mathrm{Mpc}/h$. The total mass of a truncated SIS is
\beq
M_\mathrm{SIS}=\frac{2 \sigma_\mathrm{v}^2}{G} r_\Delta.
\eeq

The total angular momentum of a halo, $J$, can be expressed in terms of a dimensionless spin parameter $\lambda$, which represents the
ratio between the actual angular velocity of the system and the hypothetical angular velocity that is needed to support the system
\citep{pee69,pad02},
\beq
J  \equiv \lambda \frac{G M^{5/2}}{|E|^{1/2}},
\eeq
where $M$ and $E$ are the total mass and the total energy of the halo, respectively. In the hypothesis of initial angular momentum acquired from tidal torquing, typical values of $\lambda$ can be obtained from the relation between energy and virial radius and the details of the spherical top-hat model \citep{pad02}. The total angular momentum of a truncated SIS can be written as
\beq
J_\mathrm{SIS}= \lambda \frac{4 \sigma_\mathrm{v}^3
r_\Delta^2}{G}.
\eeq
In general, the angular velocity $\omega$ of a halo is not constant and a differential rotation should be considered \citep{cap+al03}. However, assuming a detailed rotation pattern does not affect significantly the results. In what follows, we will consider the case of constant angular velocity. Then, $\omega$ has to be interpreted as an effective angular velocity, $\omega \simeq J_{\rm SIS}/I_{\rm SIS}$, where $I_{\rm SIS}$ is the central momentum of inertia of a truncated SIS, $I_{\rm SIS} = (2/9) M_{\rm SIS} r_\Delta^2$. In terms of the spin parameter,
\beq
\omega = 9 \lambda \frac{\sigma_\mathrm{v}}{r_\Delta}.
\eeq

Translational or rotational motions of the deflector affect its lensing properties in very different ways \citep{ser05b}. The effect due to a translational motion is a consequence of the local Lorentz invariance applied on the standard gravitoelectric field \citep{fri03,ser05b}. A peculiar velocity with respect to the local Hubble flow affects the lensing quantities through an overall multiplicative scaling factor. For slow motions, the factor takes the form $(1-v_\mathrm{los}/c)$ where $v_\mathrm{los}$ is the component of the peculiar velocity along the line of sight and $c$ the speed of light in the vacuum \citep{fri03,ser05b}; $v_\mathrm{los}$ is taken to be negative for receding lenses, i.e. for peculiar motions directed far away from the observer and towards the source. 

The problem of light deflection by a lens with angular momentum is very different in nature, since it is related to the dragging of inertial frames. The lensing effect of a spin depends on the details of the rotational motion \citep{ser02}. Gravitational lensing by a rotating isothermal sphere have been discussed in \cite{se+ca02,ser05a}. All of the lensing properties can be derived by the projected deflection potential, $\psi$. For a SIS in rigid motion,
 \beq
\label{sis11}
\psi^{\rm SIS} \simeq \left( 1-\frac{v_\mathrm{los}}{c} \right) x -L \left( \frac{3}{2}x_\Delta -x \right) x_1 ;
\eeq
lengths in the lens plane $x_1$-$x_2$ are in units of $R_\mathrm{E}$, 
\beq
\lab{sis10}
R_{\rm E} \equiv 4 \pi \left( \frac{\sigma_\mathrm{v}}{c}\right) ^2
\frac{D_{\rm d} D_{\rm ds}}{D_{\rm s}}.
\eeq
where $D_\mathrm{d}$, $D_\mathrm{s}$ and $D_\mathrm{ds}$ are the angular diameter distances from the observer to the deflector, from the observer to the source and from the deflector to the source, respectively; the dimensionless virial radius is $x_\Delta =r_\Delta /R_{\rm E}$. Equation~(\ref{sis11}) holds when the angular momentum is directed along the $x_2$-axis. The dimensionless parameter $L \equiv (2/3)(\omega R_\mathrm{E}/c)$ is an estimate of the rotational velocity. When $L>0$, the angular momentum of the lens is positively oriented along $\hat{x}_2$. The peculiar motion acts as a correction independent of the position in the lens plane. On the other hand, there are two main contributions to the gravitomagnetic correction \citep{se+ca02}. The first contribution comes from the projected momentum of inertia inside the radius $x$; the second contribution is due to the mass outside $x$ and can become significant in the case of a very extended lens, i.e. for a very large cut-off radius. We remark that the global factor $(1-v_\mathrm{los}/c)$ should apply overall, but in Eq.~(\ref{sis11}) we have neglected the higher order contribution due to its application to the gravitomagnetic term.

\section{Peculiar motion}
\label{sec:pecu}

The velocity field of galaxy clusters is a result of gravitational interaction of inhomogeneities in the large-scale mass distribution of the universe. The probability distribution  function of cluster peculiar velocities provides a tool for distinguishing between different cosmological models with differences showing up most at the high-velocity end \citep{ba+oh96}. Apart from the dependence on the cosmological density parameters, velocities scale in proportion to the normalisation constant of the matter power spectrum, which can be expressed in terms of $\sigma_8$, the rms mass fluctuation in a sphere of radius $8 h^{-1}\mathrm{Mpc}$. This parameter must then be set by requiring that the cosmological models reproduce the observed abundance of rich clusters \citep{col+al00}. For a flat $\Lambda$CDM model with $\Omega_\mathrm{M}=0.3$, $\sigma_8 =0.90$ and $h=0.7$, the three-dimensional velocity dispersion for clusters is $\gs  340~\mathrm{km~s}^{-1}$ \citep{col+al00}. It is worth noticing that the distribution of peculiar velocities for peaks of the smoothed initial density field, which can be conveniently associated with clusters, is independent of peak height \citep{col+al00}.

Peculiar velocities should decay in low-$\Omega_\mathrm{M}$ models \citep{pee93}. However, due to non-linear effects, the late time-growth of peculiar velocities is systematically underestimated by linear theory. Deviations are especially important for members of super-clusters whose velocities are about $20$ to $30\%$ larger than those of isolated clusters \citep{col+al00}.

Standard methods for determining radial peculiar velocities compare the velocity determined from the redshift with that expected for the uniform Hubble flow, $H_0 D$, where the distance to the cluster is typically determined with an empirical relationship based on Tully-Fisher (TF) or $D$-$\sigma$ distance indicators. Recently, \citet{mas+al06} calibrated the TF template with a sample of 807 galaxies in the fields of 31 nearby clusters and groups. Based on a subsample of 486 bona fide cluster members, they found a cluster velocity dispersion of $ 298 \pm 34~\mathrm{km~s}^{-1}$, in remarkable agreement with theoretical expectations. The largest peculiar velocities were found to exceed $600~\mathrm{km~s}^{-1}$. Similar results were also obtained by the POTENT program aimed to recover the three-dimensional velocity field using the expected irrotationality of gravitational flows in the weakly nonlinear regime \citep{be+de89,dek+al99}.

The bulk peculiar velocity of the cluster gas can be also measured through the kinematic component of the Sunyaev-Zeldovich effect (SZE), i.e. the change in the Cosmic Microwave Background (CMB) intensity caused by scattering \citep{su+ze80,rep95,hol+al97}. This kinematic effect appears as an increment or decrement in the CMB intensity at all frequencies. Unfortunately, actual observational uncertainties are too large to allow reliable estimates, and only limits to the bulk flow of the intermediate-redshift universe in the direction of the CMB dipole can be obtained \citep{ben+al03}.

Given the overall scaling induced by peculiar velocities on all lensing observables, the relative error in the mass estimate made when the motion along the line of sight is neglected is
\beq
\frac{\Delta M}{M} \simeq -\frac{v_\mathrm{los}}{c} .
\eeq
Observations and theoretical predictions on the velocity field discussed above suggest that the systematic error is as large as $\sim 0.3 -0.4\%$. Assuming a Gaussian velocity distribution, the effect is $\gs 0.1\%$ in nearly one third of the systems.

Whereas corrections on a single estimate at the level of the percent can not affect in a significant way complete statistical analyses, one might wonder if very deep observations of galaxy clusters could allow a detection of the gravitational lensing kinematic effect. Since the translational motion acts as an overall multiplicative factor, there is a full degeneracy between the effect of the peculiar velocity and a re-scaling of the central mass density of the cluster. Then, even using fiducial gravitational lensing data we can not disentangle such a degeneracy. 

A possible way to study the kinematic translational effect could be through joint analyses with independent data-sets. Cross-correlations of SZE surveys with lensing data should amplify the effect \citep{sc+ba06}. Future all-sky submillimetric telescopes, such as the Planck satellite\footnote{http://www.rssd.esa.int/index.php?project=Planck}, will measure the thermal SZE in many thousands of galaxy clusters. However, the smaller kinetic SZE should be detected in just few dozens. Then, for contemporary and near future lensing surveys, the kinetic correction is not supposed to play a significant role. This is also the case considering the weak-lensing power spectrum \citep{sc+ba06}.

\section{Rotating clusters}
\label{sec:spin}

Angular momentum should be presumably acquired by halos (dark matter plus gas) through tidal interactions with neighbouring objects \citep{pee69,dor70,whi84,bul+al01}. Tidal forces are stronger in dense environments, leading to more coherent rotation. Here, we are interested in coherent rotation, whereas shear flows which imply higher order gravitomagnetic effects are not considered. Recent large $N$-body simulations have given a detailed picture of the spin distribution of massive halos \citep{bet+al07,go+ye07}. The trend of the spin with the halo mass is very weak and shows a large dispersion but more-massive halos seem to have a slightly less coherent rotation in the median. The spin for massive clusters is nearly independent of the halo shape. The distribution of spins, as obtained from independent groups, can be approximated either by a log-normal distribution \citep{vit+al02,go+ye07} or by function with a longer tail at low $\lambda$ \citep{bet+al07} but anyway the main features of the distributions are pretty similar with a median value of $\lambda_\mathrm{med} \sim 0.03$ and a width of $\sigma_\mathrm{lg} \sim 0.2$ \citep{bet+al07}. The number of clusters with $\lambda \gs 0.1$ is $\sim 2\%$.

Direct observations of rotating galaxy clusters are much more uncertain. From a survey-level substructure analysis of 25 low richness clusters of galaxies contained in the 2dFGRS (2dF Galaxy Redshift Survey) cluster catalogue, \citet{bur+al04} found that 3 clusters exhibit velocity-position characteristics consistent with the presence of possible rotation, shear or infall dynamics. Recently, \citet{hw+le07} searched for rotating clusters in SDSS (Sloan Digital Sky Survey) and 2dFGRS. Out of a sample of 56 galaxy clusters with enough galaxy members with known radial velocity, they selected 6 likely rotating ones. The estimated rotation amplitudes are in the range $190~\mathrm{km~s}^{-1} \ls v_\mathrm{rot} \ls 450~\mathrm{km~s}^{-1}$ whereas the tentative velocity gradients are in the range $400~\mathrm{km~s}^{-1}\mathrm{Mpc}^{-1} \ls d v/d R \ls 800~\mathrm{km~s}^{-1}\mathrm{Mpc}^{-1}$. Even if the sample of clusters is not statistical complete, more than $10\%$ of the analyzed clusters show signatures of a rotation pattern. The ranges in velocity extend to higher values when other 6 less likely rotating clusters are included in the subsample. 
 Maybe the best case for a rotating cluster is Abell 2107, with an estimated angular velocity for the entire cluster of $d v/d R \sim 718 ~\mathrm{km~s}^{-1}\mathrm{Mpc}^{-1}$ \citep[and references therein]{kal05}.

Evidence of cluster rotation from  X-ray analyses of the intra-cluster medium is less conclusive. In principle, the presence of gas bulk velocities can be detected through Doppler shifts of X-ray spectral lines. So far, ASCA \citep{du+br05} or Chandra \citep{du+br06} observations have shown evidence for velocity gradients consistent with transitory and/or rotational bulk motion in a very few clusters. Interpreting the velocity difference for regions opposed to the centre  as due to residual gas circulation, \citet{du+br06} estimated a corresponding circular velocity of $\sim (1.2 \pm  0.7)\times 10^3~\mathrm{km~s}^{-1}$ in the Centaurus cluster. It is worth noticing that some recent numerical simulations have shown that the gas spin is $\sim 1.4$ times larger than the spin of dark matter with a tendency to decrease with halo mass \citep{go+ye07}.

The angular velocity of a cluster can be expressed in term of the spin parameter and of the overdensity as
\begin{eqnarray}
\omega & = &\frac{9}{2}\lambda H(z) \sqrt{\Delta}  \label{spi1} \\
& \simeq & 260 \left( \frac{\lambda}{0.04}\right) h~\mathrm{km~s}^{-1}\mathrm{Mpc}^{-1} \label{spi2}
\end{eqnarray}
where in Eq.~(\ref{spi2}) I have substituted for some reference values, i.e. a galaxy cluster at $z_\mathrm{d} \simeq 0.3$ with a virial overdensity of $\Delta \simeq 155.5$. For average values of the spin, the angular velocities predicted in Eq.~(\ref{spi2}) are smaller than the measurements discussed above. This can be explained if rotation is more likely detected in clusters with large spin ($\lambda \sim 0.1$). The dimensionless parameter $L$ can be written as
\begin{eqnarray}
L & = & 3 \lambda \frac{R_\mathrm{E}}{c/H(z)} \sqrt{\Delta} \label{spi3}  \\
& \simeq & 2.4 \times 10^{-5} \left( \frac{\lambda}{0.04} \right) \left( \frac{\sigma_\mathrm{v}}{800~\mathrm{km~s}^{-1} } \right)^2 ,  \label{spi4}
\end{eqnarray}
where in the second line I have considered a galaxy cluster at $z_\mathrm{d} \simeq 0.3$ with a virial overdensity of $\Delta \simeq 155.5$ and a background source population at $z_\mathrm{s} \sim 1.5$. Since spin effects are proportional to $L$, we expect them to be small.

\subsection{Strong lensing}

Detection of giant luminous arcs in the inner regions of galaxy clusters provides a tool for one of the most direct and reliable mass estimate of the inner regions. If the cluster is not far from spherical symmetry, then at first order,
\beq
\label{spi5}
M ( < \theta_\mathrm{arc}) \simeq \Sigma_\mathrm{cr}\pi ( D_\mathrm{d}\theta_\mathrm{arc} )^2,
\eeq
where $\theta_\mathrm{arc}$ is the angular radius of the arc and the mean density inside the Einstein radius equals the critical surface density $\Sigma_\mathrm{cr} =c^2 D_\mathrm{s}/(4\pi G D_\mathrm{d} D_\mathrm{ds})$. Due to lens spin, the critical curve is slightly shifted by $ \Delta \theta / \theta_\mathrm{arc} \sim L $ \citep{ser05a}. Then ignoring spin contribution affects the mass estimates by
\beq
\label{spi6}
\frac{\Delta M}{M} \simeq 2 L .
\eeq
The effect is small even for very massive and highly spinning clusters. As can be seen from Eq.~(\ref{spi4}), the relative error is $\ls 0.005\%$ for typical values of $\sigma_\mathrm{v} \sim 800~\mathrm{km~s}^{-1}$ and $\lambda \sim 0.04$ and it can be as large as $ 0.04\%$ for $\sigma_\mathrm{v} \sim 1500~\mathrm{km~s}^{-1}$ and $\lambda \sim 0.1$.

Differently from the translational motion, the rotation can imprint peculiar lensing signatures which allow in principle to distinguish the gravitomagnetic effect from that of other mass perturbations, such as a quadrupole moment \citep{ser05a}. Despite the relative variation in lensing quantities is small, the absolute variation due to the spin can be of interest. Giant luminous arcs usually form at a radial distance of $\sim 30~\mathrm{arcsec}$. Even a very tiny relative deviation of $\ls 0.01\%$ brings about a correction to the deflection angle of $\sim 3~\mathrm{milliarcsec}$, at the level of the astrometric resolution obtained with ground-based optical interferometry. This could be interesting but the real observational shortcoming is due to intrinsic size of the lensed source. In fact, either the width of thin arcs or the size of images in multiple-systems are larger than the astronometric shift due to the kinetic effect.

\subsection{Weak lensing}

In the outer regions of galaxy clusters, the deflection is small and the shear, i.e. the anisotropic distortion field, produces a weak alignment of background images, which can be detected by averaging over many near images \citep{ba+sc01}. For an axially symmetric mass distribution, images are tangentially oriented relative to the direction towards the mass centre. Rotation affects the shear components. The tangential shear corresponding to the potential in Eq.~(\ref{sis11}) is
\beq
\label{spi7}
\gamma_\mathrm{t} \simeq \frac{1}{2 x} \left( 1 - L x \sin \varphi \right)
\eeq
with $\varphi$ the polar angle in the lens plane. Then, the angular momentum of the lens gives rise to a modulation in the tangential shear which varies as the sine of the polar angle. For a rigid rotation, the amplitude of the signal ($\sim L/2$) is constant with the radius.

The relative systematic error made neglecting the rotation is $\ls L x$. In terms of the spin parameter, the uncertainty on the mass can be written as
\begin{eqnarray}
\frac{\Delta M}{M} & \ls    &  6 \lambda \left( \frac{\sigma_\mathrm{v} }{c} \right) f_{r_\Delta}  \label{spi8}  \\
                   & \simeq &  6 \times 10^{-4} \left( \frac{\lambda}{0.04} \right) \left( \frac{\sigma_\mathrm{v}}{800~\mathrm{km~s}^{-1} } \right) f_{r_\Delta}  \label{spi9}
\end{eqnarray}
where $f_{r_\Delta}(=\langle r \rangle/{r_\Delta})$ is the mean radius of the observed region in units of the viral radius. The field of view surrounding a massive cluster ($\sigma_\mathrm{v} \sim 1500~\mathrm{km~s}^{-1}$) can be explored up to large radii ($\ls 2~\mathrm{Mpc}/h$). Then, for high spins ($\lambda \sim 0.1$), the corresponding error on the mass estimate is of order $\sim 0.3\%$. 

In principle, the typical angular modulation induced by the gravitomagnetic field provides a way to measure the angular momentum in galaxy clusters. A similar effect might be artificially detected in a static mass configuration if by mistake the assumed position of the geometrical centre of the theoretical mass model does not coincide with the actual centre of the mass distribution \citep{ser02}. However, the barycentre of a well relaxed galaxy cluster can be easily identified with several reliable pointers, such as the location of the central brightest galaxy and the peak in the X-ray emission. 

In order to asses the detectability of the effect in the weak lensing regime, the gravitomagnetic correction must be compared to the main source of statistical uncertainty, which is due to the intrinsic ellipticity of the source galaxies, $\Delta \gamma_\mathrm{t}/ \gamma_\mathrm{t} \simeq \sigma_\mathrm{e}/(\sqrt{2 N} \gamma_\mathrm{t})$, where $\sigma_\mathrm{e}$($\sim 0.2-0.3$) is the intrinsic dispersion in background galaxy ellipticity per mode and $N$ is the number of background sources. Uncertainties on the tangential shear $\ls 0.01-0.02$ are routinely obtained with ground-based observations by averaging the signal over circular annuli; the total number of annuli is usually a dozen. On the other hand, for massive clusters ($\sigma_\mathrm{v} \simeq 1500~\mathrm{km~s}^{-1}$) with high spins ($\lambda \simeq 0.1$), the modulation amplitude is $\sim 10^{-4}$, two orders of magnitude smaller than the noise. 

Let us give a closer look at the effect. A coherent rotation imprints a coherent angular pattern in the lensing signal. For a nearly constant angular velocity, the signature is constant with the radius, see Eq.~(\ref{spi7}), which further helps in attempting to detect the signal. Then, the gravitomagnetic correction, when considered in subsequent circular annuli with increasing mean radius, can be viewed as a periodic function of the polar angle with period $2\pi$. The detection of a modulation is much easier to extract than a steady signal. Since the modulation is a sine function with a minus sign, the tangential shear is enhanced in the southern part of the cluster, i.e. $\pi < \varphi < 2\pi$, and vice-versa in the north. Let us consider the tangential distortion in the four quadrants of a circular annulus. The average tangential shear signal is $1/ (x_\mathrm{max} + x_\mathrm{min})$, where $x_\mathrm{max}$ and $x_\mathrm{min}$ are the outer and inner radius of the annulus, respectively.  In the north, i.e. first and second quadrant, the average signal is suppressed by $-L/\pi$; in the south, i.e. third and fourth quadrant, the signal is enhanced by $+L/\pi$. If the shear signal is averaged over the whole annulus, the gravitomagnetic contribution is washed out for a circular mass distribution. Whenever the total amplitude variation of the gravitomagnetic signal ($\sim L$) is larger than the statistical error due to the intrinsic ellipticity, there is a clear detection of the gravitational effect of the rotation. Unfortunately, this condition is fulfilled only for surface densities of the background galaxy sources, $\rho_\mathrm{back}$, well beyond actual technological capabilities. Considering a massive cluster with a large spin whose weak lensing signal is collected over large circular sectors with inner radius of $\sim 2 R_\mathrm{E}$ (to excise the central strong lensing region) and extending up to the viral radius $r_\Delta$, the gravitomagnetic tangential shear can be detected only if $\rho_\mathrm{back} \gs 10^{3}$~galaxies per arcminute squared.

Future space-borne missions or the next generation ground-based telescopes should substantially increase the observed densities of background galaxies with respect to actual values, but not enough. As an example, the proposed SNAP mission\footnote{http://snap.lbl.gov/} should get $\rho_\mathrm{back} \sim  10^{2}$~galaxies per arcminute squared, well below the requirements for the gravitomagnetic detection.

\section{Conclusions}
\label{sec:conc}

Kinematics affects mass measurements based on gravitational lensing. In order to give a quantitative estimate, I have considered as lens model a singular isothermal sphere in rigid rotation and in translational motion with respect to the background. In fact, increasing the accuracy either by considering a rotational velocity dependent on radius or a mass density profile predicted by numerical simulations would not affect results in a sensible way. Peculiar motions or coherent rotations act very differently as regards gravitational lensing but systematic deviations turn out to be $\ls 1\%$, well below actual statistical uncertainty or projection effects. The kinematic effect should not have a sizable impact on present and near-future statistical studies on scaling relations in galaxy clusters. 

As regards the detectability of the kinematic effect in galaxy clusters in the near future, prospects are not so good. The effect of translational motion can be sizable but is degenerate with an overall mass-rescaling: gravitational lensing observations by their own can not disentangle the effect. On the other hand, angular momentum imprints a distinctive feature. Due to the axially symmetric intrinsic gravitomagnetic field induced by rotation, the tangential shear shows a angular amplitude modulation and a consequent north-south asymmetry. Unfortunately, the effect is very tiny and even very deep exposures lacks the required (very high) background source density.

\section*{Acknowledgements}
This work benefited from the careful reading and the argued criticism of the referee. The author is supported by the Swiss National Science Foundation and by the Tomalla Foundation.


\setlength{\bibhang}{2.0em}

\end{document}